# Internal gravity waves from a non-local perturbation source

**V.V.Bulatov, Y.V.Vladimirov**
**Institute for Problems in Mechanics**
**Russian Academy of Sciences**
**Pr.Vernadskogo 101-1, 117526 Moscow, Russia**
**bulatov@index-xx.ru**

*Annotation.*

*The internal gravity waves far field exited by a non-local perturbation sources was considered. A separate wave mode asymptomatic presentation was constructed, describing the wave field key features depending on the source geometry.*

*Introduction and problem definition.* Internal gravity waves excited by preset temporal and spatial distribution sources propagation problem is considered. The velocity vertical component of the established wave field o $W(t, x, y, z)$ is determined by the formula [1-5]

$$LW(t, x, y, z) = Q(t, x, y, z) \qquad (1)$$

$$L = \frac{\partial^2}{\partial t^2}\left(\frac{\partial^2}{\partial x^2} + \frac{\partial^2}{\partial y^2} + \frac{\partial^2}{\partial z^2}\right) + N^2(z)\left(\frac{\partial^2}{\partial x^2} + \frac{\partial^2}{\partial y^2}\right)$$

Brunt-Vaisala frequency $N^2(z) = -gd\ln\rho/dz$ ($g$ is gravitational acceleration, $\rho$ is stratified medium density), L is the linear internal gravity waves operator in the Boussinesq approximation. The function $Q$ describes distribution density of wave field perturbation nonlocal sources and for example in case of impulsive point source of mass $Q = \delta(x)\delta(y)\delta'(z-z_0)\delta''(t)$, for impulsive normal force $Q = \delta(t)\delta(z-z_0)(\delta''(x)\delta(y) + \delta''(y)\delta(x))$ [1-9]. Let us in all cases assume that the sources $Q$ is enabled with $t \geq 0$, and with $t < 0$ medium is at a standstill, i.e.: $W \equiv 0$ with $t < 0$. It is obvious that if wave oscillations in total depth stratified layer are considered $-H < z < 0$, additional boundary conditions should be made, for example of "hard cover".: $G = 0, z = 0, -H$ [1-5]. By reason of the problem linearity it is sufficient to consider Green's function, i.e. the solution $G(t, x, y, z, z_0)$ of the equation: $LG = \delta(x)\delta(y)\delta(z - z_0)\delta(t)$ ($z_0$ is the point source depth), which represents modes total: $G = \sum_n G_n$.

The problem (1) solution with corresponding boundary and initial conditions represent triple quadratures total, which significantly complicates both the function $W(t, x, y, z)$ numerical calculation and its qualitative assay [6-9]. However as one may expect that the nonlocal source exact form will nave no affect on internal waves field at some distance of it, if this source width is less than internal gravity characteristic lengths and therefore for example it is reputed, that perturbation nonlocal source transverse section is described by a fairly simple function and numerical calculations made for various functions show that indeed internal gravity waves field weakly depends on particular selection of perturbation nonlocal source transverse section [1-5].

It is clear in the exact computational solution of the problem (1) that as a minimum three various regions of internal waves established field [6-9]. First this is the region directly over the nonlocal source, which is in the order of medium layer across – near zone, in which internal gravity waves wave field in near zone weakly depends on particular stratification and amplitude of velocities and elevations are maximum. At big distances of nonlocal sources the internal gravity waves field falls into separate modes while each mode is enclosed in its Mach wedge, outside wedge amplitude is small, besides at big distances in case of weak stratification streamlined nonlocal source exact form can be ignored and substituted with corresponding source and drain system [1-9]. Such investigative techniques of internal gravitation waves can be applied to only relatively few simple form streamlined bodies (sphere, cylinder, ovoid etc.).

Therefore far from the perturbation streamlined nonlocal source it is reasonable to substitute source distribution described by this nonlocal source form with a simpler combination of sources and drains [1-5], for example to conceive approximately the function $Q$ of as: $Q = Q_0 \delta(y)(\delta(x_+) - \delta(x_-))$, where $x_+$, $x_-$ are the source and drain coordinates accordingly, $Q_0$ is a constant. Using the asymptomatic representation $G_n$ far from perturbation sources it is possible to get the expression for a separate mode $W_n$ of the complete field W as: $W_n \approx Q_0(G_n(x_+, y, z, z_0) - G_n(x_-, y, z, z_0))$. Substitution of the perturbation streamlined nonlocal source with a corresponding system of sources and drains as well as use of Green's function asymptomatic representations allows to compute internal gravitation field without recourse to tedious numerical calculations of corresponding reiterated quadratures.

***Nonlocal sources far fields.*** Let us later consider internal gravitation waves far field generated by the of horizontally distributed sources system, with the aid of which it is possible to solve the problems of mathematical modeling of nonlocal sources perturbation, longitudinal size of which are much more than transverse sizes. Let us denote weight function by p($\xi$) labeling sources distribution lateral density while we assume $\int_{-\infty}^{\infty} p(\xi)\, d\xi = 1$. By reason of the problem linearity the total field of for example elevation $\eta$ ($W = \frac{\partial \eta}{\partial t}$) is superposition of field of point sources having a capacity of

$$\eta(\xi, y, z, z_0) = \int_{-\infty}^{\infty} G(\xi - u, y, z, z_0)\, p(u)\, du \qquad (2)$$

where $\xi = x + Vt$, $G(\xi, y, z, z_0)$ is the Green's function for a perturbation source moving at a rate of $V$, various asymptomatic representations for this function were obtained in [1-5]. If the source typical dimensions are comparable to the Green's function G first pulse width, the

convolution( 2 ) gives maximum values of internal gravity waves elevation field η in wave front set adjacency and, while oscillating, with wave front set offset rapidly decays as far form wave field the $G$ function oscillation frequency increases and neighboring half-waves dampen each other.

With frontal distance y increase the function G first half-wave length increases (proportionally to $y^{1/3}$ and distributed source field becomes more and more similar to point source field. Let us check this qualitative reasoning on the following model situation, within the framework of which let us compute an integral ( 2 ).

Let us take wave field representation in terms of Airy waves as the function $G$ which describes asypmtomatics far from perturbation point source moving at a rate of $V$ [1-5 ]

$$G(\xi, y, z, z_0) = \sum_{n=1}^{\infty} \frac{V\alpha_n^2 f_{n0}(z) f_{n0}'(z_0)}{2(3\beta_n \xi_n)^{1/3}} Ai\left(\frac{\alpha_n \xi - y}{(3\beta_n \xi_n)^{1/3}}\right),$$

where $Ai(x)$ is Airy function, $\xi = x + Vt$, $\xi_n = y/\alpha_n$, eigenfunctions $f_{n0}(z)$ and eigenvalues $\alpha_n$ are defined from the spectral problem [1-5]

$$\frac{d^2 f_{n0}(z)}{dz^2} + \frac{\alpha_n^2 + 1}{\alpha_n^2 V^2} N^2(z) f_{n0}(z) = 0$$

$$f_{n0}(0) = f_{n0}(-H) = 0. \qquad \int_{-H}^{0} f_{n0}^2(z) N^2(z) dz = 1$$

$$\beta_n = a_n \lambda_n^5 V^4 \alpha_n^5 = \frac{a_n V^4}{(V^2 - c_n^2)^{5/2}}$$

Made numerical calculations show that in case of a source having some typical longitudinal size less than Airy wave typical wave pulse width the function $\eta(\xi, y, z, z_0)$ weakly depends on distribution specific form $p(\xi)$, therefore let us later consider unit intensity

Gauss distribution $p(\xi) = \frac{1}{\sqrt{2\pi}} \exp(-\frac{\xi^2}{2\sigma^2})$, where root-mean-square deviation $\sigma$ is assumed as perturbation nonlocal source typical lateral size.

Within this framework of these two assumptions there's no loss of generality while the integral (2) is successful in analytical calculation. By employing the Airy function integral representation: $Ai(x) = \frac{1}{2\pi} \int_{-\infty}^{\infty} \exp(i(\frac{z^3}{3} - xz)) dz$ the formula (2) can be written as

$$\eta_n(\xi, y, z, z_0) = \frac{B_n(z, z_0)}{(2\pi)^{3/2} \sigma} \int_{-\infty}^{\infty} \int_{-\infty}^{\infty} \exp(i(\frac{\tau^3}{3} - \frac{u - \xi_n}{D_n}\tau) - \frac{(\xi - u)^2}{2\sigma^2}) d\tau\, du$$

$$B_n(z, z_0) = \frac{V \alpha_n^2 f_{n0}(z) f'_{n0}(z_0)}{2(3\beta_n \xi_n)^{1/3}}$$

where $D_n = \frac{(3\beta_n \xi_n)^{1/3}}{\alpha_n}$ is the Airy wave first pulse typical length for the nth wave mode.

Integral over $u$ can be calculated

$$\eta_n(\xi, y, z, z_0) = \frac{B_n(z, z_0)}{(2\pi)} \int_{-\infty}^{\infty} \exp(i(\frac{\tau^3}{3} - \frac{\xi - \xi_n}{D_n}\tau) - \frac{\tau^2 \sigma^2}{2 D_n^2}) d\tau$$

It is necessary to substitute variables in the resulting integral in order to eliminate quadratic in $\tau$ terms. By letting down a contour on the real axis as it is possible due to y $Jm\, u^3$ positive sign, one get obtain the answer in form of the Airy function by exponent

$$\eta_n(\xi, y, z, z_0) = B_n(z, z_0) Ai(d_n - \frac{\Delta_n^4}{4}) \exp(\frac{\Delta_n^2}{2}(-d_n + \frac{\Delta_n^4}{6})) \qquad (3)$$

where the parameters $d_n = \frac{\xi - \xi_n}{D_n}$, $\Delta_n = \frac{\sigma}{D_n}$ define accordingly the distance from the viewpoint to the front and horizontal source typical size related to the typical length of the nth mode first pulse for a point source. Let us remark that the formula (3) can be used both for field lengthwise sections ( y is fixed, ξ is variable) and for cross cuts (ξ is fixed, y is variable), in the latter case $d_n = \dfrac{\xi_n - y/\alpha_n}{D_n}$.

The Fig. 1 represents the results of the internal gravity waves elevation first mode $\eta_1$ calculation by the formula (3) without considering the factor $B_n(z, z_0)$ for a point ($\Delta = 0$, unbroken line) and horizontally distributed source ($\Delta = 1.2$, dotted line). From the figure it is seen, that for a distributed source wavelength is slightly variable, wave amplitude damps and the first peak becomes more and more dominant.

The Fig. 2 represents numerically obtained the first peak height to the second one ratio δ depending on the value $\Delta$. From the figure it is seen, that with $\Delta > 1$ the first peak becomes markedly dominant. From the figures and formulae for $d_n$ and $\Delta_n$ it also follows, that for the fixed σ with the nth mode frontal offset from the source $D_n$ increases, $\Delta_n$ decreases and the field is described by the Airy function for a point source more and more closely, which is in full conformity with above reasoning.

The Fig. 3 represents the results of calculation of the internal gravity waves vertical velocity first mode $W_1$ to the "local" Airy approximation for a point source and a distributed one (value $\Delta = 0.8$), from which it is seen, that indeed if a source typical sized are comparable to the Green's function G first pulse width, the convolution gives maximum values of internal gravity waves η field in wave front set adjacency and, while oscillating, with wave front set offset rapidly decays as far form wave field the $G$ function oscillation frequency increases and neighboring half-waves dampen each other.

Thus constructed asymptomatic representations of far wave fields make it possible to study perturbation nonlocal sources internal gravity waves generation peculiarities.

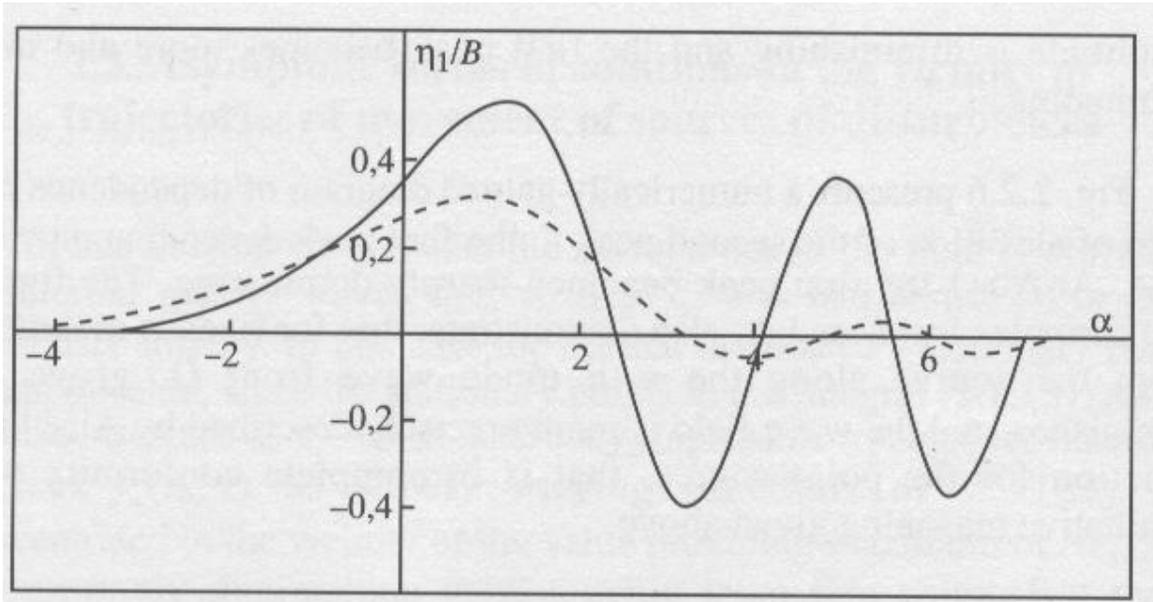

Fig.1 First mode on internal gravity wave elevation for the point type and nonlocal sources.

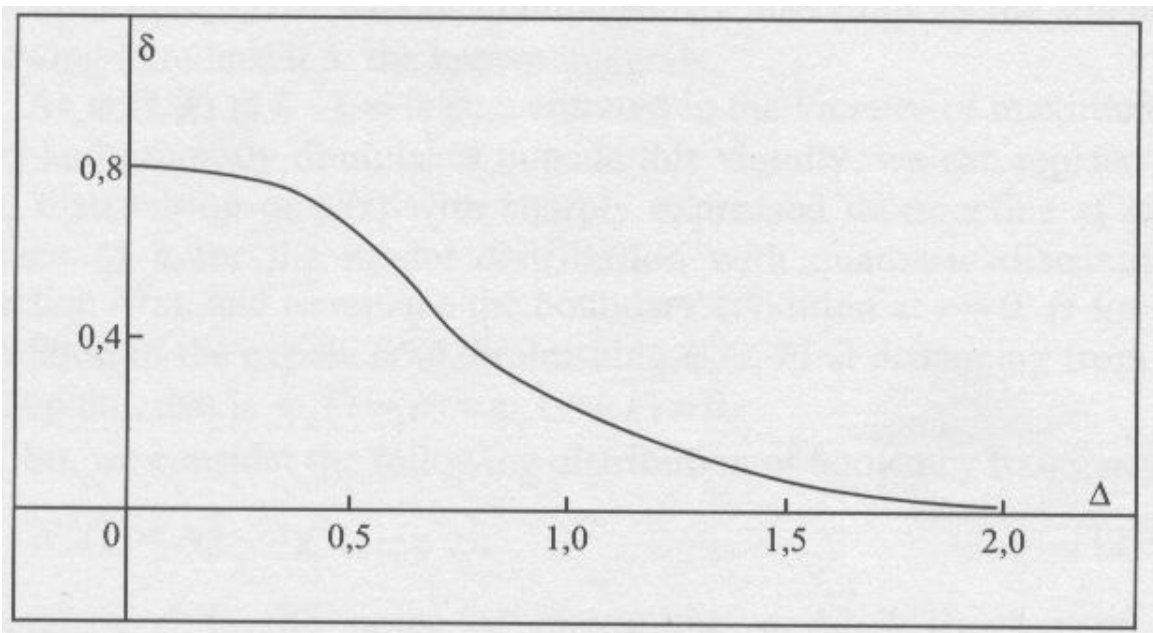

Fig.2 Relationship of the first two maximums of elevation for the nonlocal source.

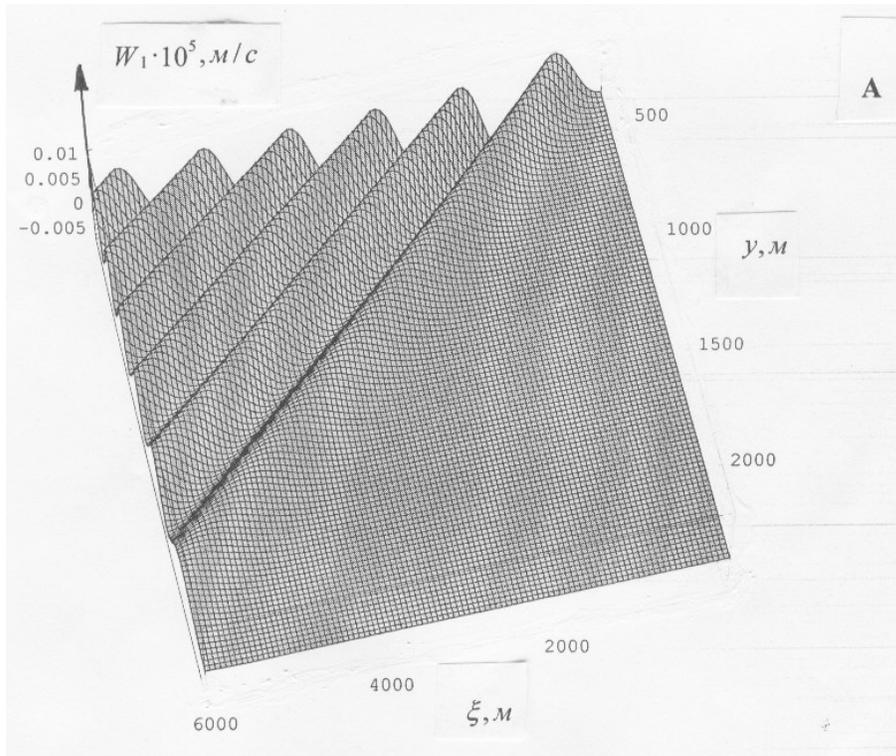

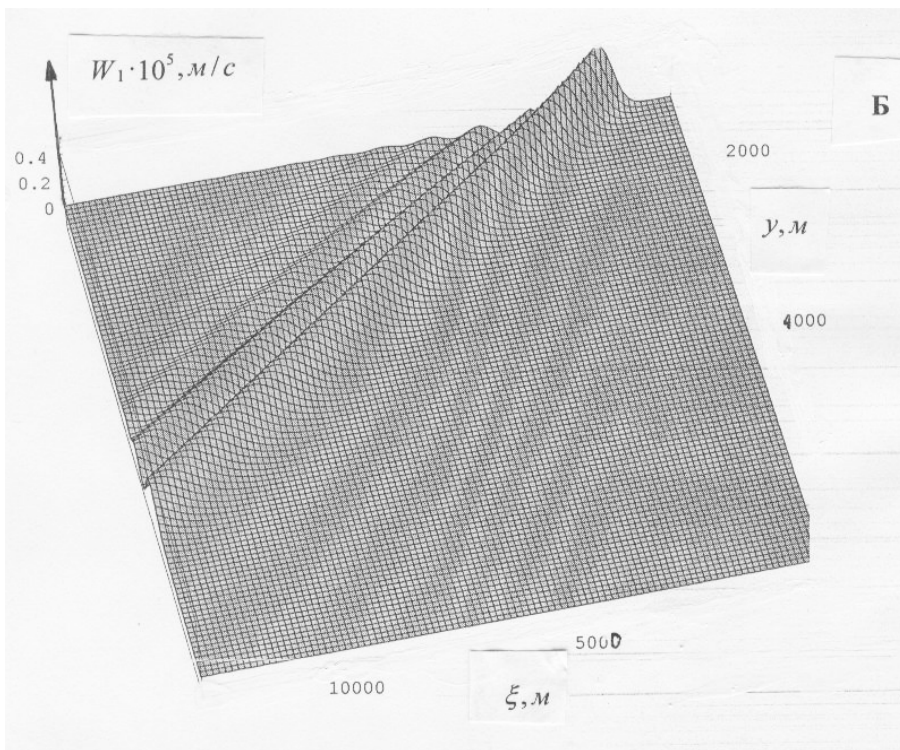

Fig.3 First mode of internal gravity wave vertical velocity: А – point perturbation source, Б – nonlocal perturbation source.